%% file: nhsub1.tex
\documentclass[%
 reprint,
nofootinbib,
nobibnotes,
 amsmath,amssymb,
 aps,
floatfix,
]{revtex4-2}

\usepackage{graphicx,color}% Include figure files
\usepackage{bm,url,float,ulem}% bold math
\usepackage{multirow}
\usepackage{booktabs}
\usepackage{caption}
\usepackage{subfigure}
\usepackage{float}
\usepackage{subfig}
\newcommand\erase{\bgroup\markoverwith{\textcolor{red}{\rule[.5ex]{2pt}{0.4pt}}}\ULon}

\begin{document}

\preprint{APS/123-QED}

\title{Quiescent luminosities of transiently accreting neutron stars with neutrino heating due to charged pion decay}% Force line breaks with \\

\author{He-Lei Liu}
\email{heleiliu@xju.edu.cn}
\affiliation{School of Physical Science and Technology, Xinjiang University, Urumqi 830046, China}
\author{Zi-Gao Dai}
\affiliation{Department of Astronomy, School of Physical Science, University of Science and Technology of China, Hefei 230026, Anhui, China}
\author{Guo-Liang L\"{u}}
\affiliation{School of Physical Science and Technology, Xinjiang University, Urumqi 830046, China
}
\author{Akira Dohi}
\affiliation{Department of Physics, Kyushu University, Fukuoka 819-0395, Japan}
\affiliation{Interdisciplinary Theoretical and Mathematical Sciences Program (iTHEMS), RIKEN, Wako 351-0198, Japan}
\author{Gao-Chan Yong}
\affiliation{Institute of Modern Physics, Chinese Academy of Sciences, Lanzhou 730000, China}
\affiliation{School of Nuclear Science and Technology, University of Chinese Academy of Sciences, Beijing, 100049, China}
\author{Masa-aki.~Hashimoto}
\affiliation{Department of Physics, Kyushu University, Fukuoka 819-0395, Japan}
%\author{Gao-Chan Yong}
%\affiliation{Institute of Modern Physics, Chinese Academy of Sciences, Lanzhou 730000, China}

\date{\today}% It is always \today, today,

\begin{abstract}
We study the quiescent luminosities of accreting neutron stars by a new mechanism as neutrino heating for additional deep crustal heating, where the neutrino heating is produced by charged pions decay from the nuclear collisions on the surface of neutron star during its active accretion. For low mass neutron star($\lesssim1.4~M_{\odot}$), as the neutrino heating is little($\lesssim1$ MeV per accreted nucleon) or there would be no neutrino heating, the quiescent luminsoity will be not affected or slightly affected. While for massive neutron star ($\gtrsim2~M_{\odot}$), the quiescent luminosity will be enhanced more obviously with neutrino heating in the range 2-6 MeV per accreted nucleon. The observations on cold neutron stars such as 1H 19605+00, SAX J1808.4-3658 can be explained with neutrino heating if a fast cooling and heavy elements surface are considered. The observations on a hot neutron star such as RX J0812.4-3114 can be explained with neutrino heating if the direct Urca process is forbidden for a massive star with light elements surface, which is different from the previous work that the hot observations should be explained with small mass neutron star and the effect of superfluidity.

\end{abstract}

%Use showkeys class option if keyword
\maketitle

%=== Name of Journals =====
\input{./Name_Journal}

%\tableofcontents

\section{Introduction}
%SXTs
Transiently accreting neutron stars in low-mass X-ray binaries accrete matter episodically from their low-mass companions during outburst~\cite{2017JApA...38...49W}. The outburst periods can last weeks to months (ordinary transients), or years to decades (quasi-persistent transients) with luminosity $\sim10^{36}-10^{38}~\rm erg~s^{-1}$. The quiescent periods can last years to decades, during which the accretion is turned off or strongly suppressed with low luminosities $10^{31}-10^{34}~\rm erg~s^{-1}$. The systems are also called soft x-ray transitents (SXTs). A comparison of the theory of quiescent luminosity of neutron star in SXTs with observations can provide important information. Since the heating and cooling depend on neutron star properties such as deep crustal heating, shallow heating, mass, equation of state, superfluidity, composition of the stellar core and outer layers, such a comparison can help us understand both the crust and core properties of neutron star~\cite{2021PhR...919....1B,2021PhRvD.103h3004F,2003A&A...407..265Y,2020MNRAS.496.5052P,2021A&A...645A.102P,2019A&A...629A..88P}.

%deep crust heating and shallow heating
Observed quiescent luminosities of transiently accreting neutron stars are generally believed to be held by deep crustal heating with the release of $\sim1-2~\rm MeV$ per accreted nucleon~\cite{1998ApJ...504L..95B,1990A&A...227..431H,2003A&A...404L..33H,2008A&A...480..459H,2018A&A...620A.105F}, which is produced by the non-equilibrium nuclear reactions such as electron captures, neutron emission/absorption and pycnonuclear reaction due to the continuous compression during outburst phase. It's worth to mention that Chugunov and collaborators have pointed out that the mobility of neutrons in the inner crust could strongly suppress the heat release and drastically change its distribution over density~\cite{2020MNRAS.495L..32C,2021PhRvD.103L1301G,2021MNRAS.507.3860S}, compared with the conventional deep crustal heating models (Hasensel et al. Refs.~\cite{1998ApJ...504L..95B,1990A&A...227..431H,2003A&A...404L..33H,2008A&A...480..459H,2018A&A...620A.105F}). There have been many works about the study of quiescent luminosities of transiently accreting neutron stars~\cite{2003A&A...407..265Y,2020MNRAS.496.5052P,2015MNRAS.447.1598B,2015MNRAS.452..540B,2017PhRvC..96c5802H,2021A&A...645A.102P,2019A&A...629A..88P,2021PhRvD.103f3009L,2018IJMPE..2750067M}. From these studies, a strong superfluidity in the cores of low mass stars and a fast neutrino emission in the cores of high-mass stars are needed to explain the observations. The theoretical models with fast neutrino emission due to nucleon direct Urca or pion condensation are proposed under the assumption that the crust and core are in thermal equilibrium, where the heating process includes the only crustal heating one for quiescent luminosity calculation. On the other hand, several observations of crust cooling after their outbursts~\cite{2017JApA...38...49W}, such as some ordinary SXTs (IGR J17480-2446, Swift J174805.3-244637, 1RXS J180408.9-342058, Aql X-1) and quasi-persistent SXTs (MAXI J0556-332, EXO 0748-676, MXB 1659-298, XTE J1701-462, KS 1731-260, HETE J1900.1-2455), indicate a need of shallow heating ($0$--$17~\rm MeV/u$) at the depth $10^{8-10}~\rm{g~cm^{-3}}$~\cite{2009ApJ...698.1020B,2015ApJ...809L..31D,2016MNRAS.456.4001W,2020PhRvC.102a5804C}. However, the heating source is still unknown~\cite{2016MNRAS.461.4400O,2018MNRAS.477.2900O,2020PhRvC.102a5804C}.
%neutrino heating

Recently, a new source as neutrino heating in neutron star deep crusts from the decay of charged pions on the neutron star surface has been proposed~\cite{2018PhRvC..98b5801F}. The charged pions are produced in nuclear collisions on the neutron star surface during active accretion. For massive neutron stars, neutrinos deposit $\sim2~\rm MeV$ of heat per accreted nucleon into the deep crust ($10^{12-13}~\rm g~cm^{-3}$). The crust cooling of MXB 1659-29 has been studied with an extra heating from neutrinos~\cite{2018PhRvC..98b5801F}. It shows that a higher thermal conductivity for the neutron star inner crust is needed to explain the observation. However, a question appears: how does the extra neutrino heating affect the quiescent luminosity of transiently accreting neutron stars? To answer this question, we investigate the quiescent luminosity of SXTs with neutrino heating in this paper.

An outline of the paper is as follows. In section \ref{sec:nh} we describe the pion production and the energy deposition by neutrinos from the stopped pion decays. Section \ref{sec:eqa} gives the basic equations of thermal evolution of neutron stars and physics input. Section \ref{sec:result} presents our results of quiescent luminosity of SXTs with neutrino heating. Finally, we offer our conclusion and outlook in Section \ref{sec:conc}.

\section{pion production and neutrino heating }
\label{sec:nh}

 A neutron star residing in a low-mass x-ray binary accretes material (typically hydrogen-rich or helium-rich matter) onto its surface from the companion in an accretion disk~\cite{2001NewAR..45..449L}. The strong gravity of neutron star accelerates incoming particles to kinetic energy before they reach the neutron star surface.

Assuming that the infalling matter is in free falling, the kinetic energy $T$ of the accreted matter on the neutron star surface is~\cite{1992ApJ...384..143B}:
\begin{eqnarray}
T=m_0c^2\left(\frac{1}{\sqrt{1-2GM/c^2R}}-1\right),
\label{eq:0}
\end{eqnarray}
where $m_0$ indicates the mass of the infalling particle. $M$ and $R$ are the mass and radius of neutron star, respectively. As a result, the kinetic energy of particles will be several hundred MeV per nucleon before they reach the neutron star surface. For example, an accreted proton will get the kinetic of 585 MeV before they reach a $2M_{\odot}$  and radius of 10 km neutron star surface, while for a $1.4M_{\odot}$ and radius of 10 km neutron star, an accreted proton will get the kinetic of 286 MeV before they get the neutron star surface. During outburst, these incoming particles collide with the nuclei on the surface of neutron star. Then, if the kinetic energy of the particles is higher than the pion production threshold ($\approx290$ MeV), nuclear collisions produce pions. Neutral pions decay instantaneously via
 \begin{eqnarray}
\pi^{0}\rightarrow \gamma+\gamma~~,
\end{eqnarray}
and the energy will be released at the stellar surface. Positively charged pions slow down and stop near the neutron star surface and decay via
\begin{equation}
\begin{aligned}
\pi^{+} &\rightarrow \mu^{+}+\nu_{\mu}~~,\\
\mu^{+} &\rightarrow e^{+}+\nu_{e}+\bar{\nu}_{\mu}~~,
\label{eq:posi}
\end{aligned}
\end{equation}
These processes produce muon neutrinos of energy $E_{\nu_{\mu}}=29.8~\rm MeV$, electron neutrinos of energy $E_{\nu_{e}}=33.3~\rm MeV$, anti-muon neutrinos of energy $E_{\bar{\nu}_{\mu}}=37.7~\rm MeV$, which are estimated from the neutrino energy spectrum~\cite{2006PhRvD..73c3005S,2018PhRvC..98b5801F}. Negative charged pions decay via the similar processes

\begin{equation}
\begin{aligned}
\pi^{-} &\rightarrow \mu^{-}+\bar{\nu}_{\mu}~~,\\
\mu^{-} &\rightarrow e^{-}+\bar{\nu}_{e}+{\nu}_{\mu}~~,
\end{aligned}
\end{equation}

Assuming half of the neutrinos produced move into the crust and the other half escape from the neutron star, the total energies were carried into the crust per accreted nucleon as follow:
\begin{equation}
\begin{aligned}
q_{\nu}&\approx 0.5(E_{\nu_{\mu}}+E_{\nu_{e}}+E_{\bar{\nu}_{\mu}})(N_{\pi^{+}}+N_{\pi^{-}})\\
&=50.4~{\rm MeV}(N_{\pi^{+}}+N_{\pi^{-}})~~,
\end{aligned}
\label{eq:nh1}
\end{equation}
where $N_{\pi^{+}}$, $N_{\pi^{-}}$ are the total numbers of $\pi^{+}$, $\pi^{-}$ produced per accreted nucleon, respectively. As the experiment data of charged pion production from the interaction of nucleus-nucleus collisions still remains incomplete, we can estimate $N_{\pi^{+}}$ and $N_{\pi^{-}}$ from the isospin-dependent Boltzmann-Uehling-Uhlenbeck (IBUU) transport simulations~\cite{2002PhRvL..88s2701L,2021PhRvC.104a4613Y,2019JPhG...46j5105Y,2017PhRvC..96d4605Y}.

The number of pion production for per collision event depends on both the initial kinetic energy and the type of the target nuclei. The former is determined by the gravitational acceleration and electromagnetic acceleration. The latter depends on the accreted matter from the companion star. As a result, many types of nuclear collisions are involved with the efficiency of neutrino heating due to pion production. For simplicity, only $\alpha-\rm Fe$ nuclear collisons are considered instead of a series of nuclear collisions in accordance with Ref.~\cite{2018PhRvC..98b5801F}. The neutrino energy ($q_{\nu}$) deposition in the inner crust can reach 2.82 MeV in $\alpha-Fe$ collision for positive charged pion decay for a $2M_{\odot}$ neutron star(see more details in Table II in Ref.~\cite{2018PhRvC..98b5801F}). For a $1.4M_{\odot}$ neutron star, on the other hand, $q_{\nu}$ due to positive charged pion decay is very small ($\sim0.1$~\rm MeV) or zero. In Ref.~\cite{2018PhRvC..98b5801F}, the negative charged pion decay is not considered because $\pi^{-}$ is easy to be absorbed. However, $\pi^{-}$ can not be absorbed absolutely, only $18\%$ of the charged pions are absorbed for Au+Au collision at a beam energy of 400 MeV/nucleon~\cite{2020arXiv200902674G}.  The ratio of $\pi^{-}/{\pi^{+}}$ is in the range $\sim 1-3$ from both experiment~\cite{2010NuPhA.848..366R} and theory simulations~\cite{2015PhRvC..91d4609L,2016PhRvC..94f4621C,2017PhRvC..96d4605Y,2019JPhG...46j5105Y,2021PhRvC.104a4613Y}. As a result, the neutrino energy from both $\pi^{+}$ and $\pi^{-}$ decay should be considered.

\begin{figure}[htbp]
    \centering
    \includegraphics[width=0.7\linewidth,angle=-90]{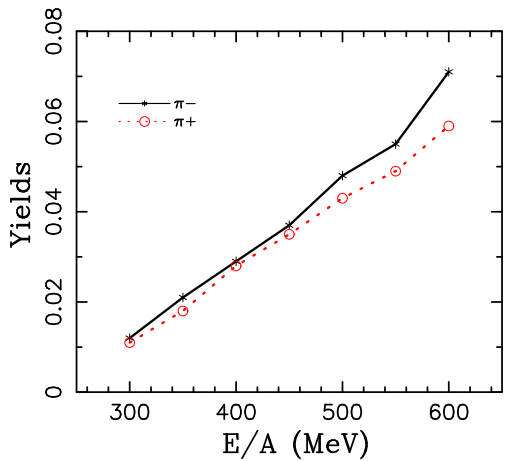}
    \caption{The pion production as a function of beam energy with a random impact parameter b= 0-6.1 fm for $\alpha$-Fe collisions estimated using the IBUU transport model.}
    \label{fig:pion}
\end{figure}

FIG.~\ref{fig:pion} shows the pion production as a function of beam energy for $\alpha$-Fe collisions, which is estimated by using the IBUU transport model. For a detailed description of this transport model, one can found in the Refs.~\cite{2002PhRvL..88s2701L,2017PhRvC..96d4605Y}.
The results are obtained by running 25000 events at each incident beam energy.  The yields of pion production, represent the number of pions produced per collision event.
One can see the values of $N_{\pi_{+}}$ and $N_{\pi_{-}}$ more detail in Table \ref{tab:pionp}. Meanwhile, the values of $Q_{\nu}$ at different incident beam energy are given in Table \ref{tab:pionp} based on Eq. (\ref{eq:nh1}). It is worth noting that, $Q_{\nu}=1.16~\rm MeV$ at the incident beam energy of 300 MeV per nucleon, $Q_{\nu}=6.55~\rm MeV$ at the incident beam energy of 600 MeV per nucleon.

\begin{table}[tbp]
\caption{The beam energy, charged pion production ($N_{\pi_{+}}$, $N_{\pi_{-}}$) and neutrino energy $q_{\nu}$ for $\alpha$--${\rm Fe}$ collision.}
\label{tab:pionp}
\centering
\begin{tabular}{cccc}
\hline\hline
E/A (\rm MeV)  & $N_{\pi^{-}}$ & $N_{\pi^{+}}$ & $q_{\nu} (\rm MeV)$  \\
\hline
300 & 0.012  & 0.011 & 1.16  \\
\hline
350 & 0.021 & 0.018  & 1.97  \\
\hline
400 & 0.029 & 0.028 & 2.87 \\
\hline
450 & 0.037 & 0.035 & 3.63 \\
\hline
500 & 0.048 & 0.043 & 4.59 \\
\hline
550 & 0.055 & 0.049 & 5.24 \\
\hline
600 & 0.071 & 0.059 & 6.55 \\
\hline\hline
\end{tabular}
\end{table}

Besides the possible absorption of charged pions, the accreted matter arrives at the neutron star surface with close to the free-fall velocity under the condition that the innermost stable orbit lies outside the neutron star or the accretion disk is truncated by the magnetic field of neutron star, if the fall starts from the inner edge of the accretion disk, the kinetic energy will be less than the value estimated from Eq.~(\ref{eq:0}). Also, the free-fall of the matter can be decelerated by radiative pressure from the surface heated by the accretion, which would decrease the particle's kinetic energy. As a result, the efficiency of pion would be reduced. However, the accreting particles may undergo electromagnetic acceleration in the strong electric and magnetic fields which would enlarge the particle's kinetic energy, the total number of charged pions could be increased. Thus, in this work, we adopt $q_{\nu}=0-1~\rm MeV$ per accreted nucleon for $1.4M_{\odot}$ neutron star while $q_{\nu}=0-6~\rm MeV$ per accreted nucleon for $2M_{\odot}$ neutron star in the quiescent luminosity calculations.

\section{Basic equations and physics input}\label{sec:eqa}

\subsection{Basic equations}
The spherically symmetric general relativistic stellar evolutionary code firstly developed by Ref.~\cite{1984ApJ...278..813F} was used for our quiescent luminosity calculations of accreting neutron stars, the basic equations in hydrostatic equilibrium are formulated as follows~\cite{1977ApJ...212..825T}:
\begin{equation}
    \frac{\partial M_{tr}}{\partial r}=4\pi r^2\rho,
 \label{eq:1}
\end{equation}

\begin{equation}
%\begin{split}
    \frac{\partial P}{\partial r}=-\frac{GM_{tr}\rho}{r^2}\left(1+\frac{P}{\rho \rm c^2}\right)\left(1+\frac{4\pi r^3P}{M_{tr}\rm c^2}\right)V^{2},
%    \end{split}
\label{eq:2}
\end{equation}

\begin{equation}
    \frac{\partial \left(L_{r}e^{2\phi/c^2}\right)}{\partial M_r}=e^{2\phi/c^2}\left(\varepsilon_n-\varepsilon_\nu+\varepsilon_g\right)~~,
\label{eq:3}
\end{equation}

\begin{equation}
\begin{aligned}
   \frac{\partial {\rm ln} T}{\partial {\rm ln} P}=&\frac{3}{16\pi acG}\frac{P}{T^{4}}\frac{\kappa L_{r}}{M_{r}}\frac{\rho_{0}}{\rho}\left(1+\frac{P}{\rho c^{2}}\right)^{-1}\left(1+\frac{4\pi r^{3}P}{M_{tr}c^{2}}\right)^{-1}\\
   &\times\left(1-\frac{2GM_{tr}}{rc^{2}}\right)^{1/2}+\left(1-\left(1+\frac{P}{\rho c^{2}}\right)^{-1}\right),
\end{aligned}
\label{eq:4}
\end{equation}

\begin{equation}
    \frac{\partial \phi}{\partial M_{tr}}=\frac{G\left(M_{tr}+4\pi r^3 P/\rm c^2\right)}{4\pi r^4\rho}V^2,
\label{eq:6}
\end{equation}
where
\begin{equation}
 \frac{\partial M_{tr}}{\partial M_{r}}=\frac{\rho}{\rho_{0}}V^{-1}~~,
V\equiv\left(1-\frac{2GM_{tr}}{r\rm c^2}\right)^{-1}.
\end{equation}
The quantities used in the above equations are defined as follows:
$M_{tr}$ and $M_r$ are the gravitational and rest masses inside a sphere of radius $r$, respectively; $\rho$ and $\rho_0$ are the total mass energy density and rest mass density, respectively; $P$ and $T$  indicate the local pressure and temperature, respectively;  $\varepsilon _{n}$ is the heating rate by nuclear burning, $\varepsilon_{\nu}$ denote neutrino energy loss, $\varepsilon_{g}$ is the gravitational energy release; $c$ and $G$ are the light velocity and gravitational constant, respectively. $a$ is the Stefan-Boltzmann constant; $\phi$ is the gravitational potential in unit mass, $\kappa$ is the opacity which is calculated by using the public conductivity codes~\cite{2001A&A...374..151B,2015SSRv..191..239P}.

\subsection{Physics input}

%EOSs
For the equation of state (EOS), we adopt Togashi and BSK24 EOSs which are widely used to simulate various astrophysical phenomena~\cite{2017NuPhA.961...78T,2019PTEP.2019k3E01D,2018A&A...620A.105F,2013PhRvC..88b4308G,2019A&A...629A..88P}. As the Togashi EOS has the low symmetry energy, the nucleon direct Urca process is prohibited with such an EOS. To explain some cold NS observations, another exotic cooling model due to $\pi$ condensation has been proposed~\cite{2021PhRvD.103f3009L,2021AD}. As a result, we also adopt Togashi+$\pi$ EOS for a comparison.
The kinetic energy of the incoming particles which depend on the compactness of neutron star strongly affect the pion production, the compactness of neutron star is sensitive to the equation of state. We show the incoming kinetic energies of a proton at the surface of neutron star predicted  by the three representative equations of state in table \ref{tab:KE}.
\begin{table}[tbp]
\caption{The radii $R$ of a 1.4 and 2.0$M_{\odot}$ neutron stars as well as the incoming kinetic energies of a nucleon ($m_N=939~\rm MeV$) at the surface of neutron star predicted by the three representative equations of state discussed in the text.}
\label{tab:KE}
\centering
\begin{tabular}{ccccc}
\hline\hline
Model & $R_{14}~(\rm km)$ & $T_{14}~\rm(MeV)$ & $R_{20}~\rm(km)$ & $T_{20}~\rm(MeV)$ \\
\hline
Togashi & 11.55 & 232.1 & 11.17  & 427.2 \\
\hline
Togashi+$\pi$& 10.97 & 249.7 & 10.33 & 493.8 \\
\hline
BSK24& 12.54 & 207.2 & 12.27& 363.5\\
\hline\hline
\end{tabular}
\end{table}
As there are many uncertainties on pion production(e.g., electromagnetic acceleration on the kinetic energy of the incoming particle, the radiative pressure deceleration on the kinetic energy of the incoming particle, the nuclei on the surface of neutron star), we can give an approximate vale of $q_{\nu}=0-1 ~\rm MeV$ per accreted nucleon for $1.4M_{\odot}$ and $q_{\nu}=0-6 ~\rm MeV$ per accreted nucleon for $2.0M_{\odot}$ with the above three equation of states.

%deep crustal heating
The heating rate $\varepsilon _{n}$ includes the standard deep crustal heating and the extra neutrino heating in the inner crust ($10^{12}-10^{13}~\rm g~cm^{-3}$). As a result, the deep crustal heating has the following form:
\begin{equation}
    Q_i=6.03\times\dot{M}_{-10}(q_{i}+q_{\nu_i})10^{33}\ \rm erg\ s^{-1},
\label{eq:ch}
\end{equation}
where $q_i$ and $q_{\nu_i}$ are the effective heat per nucleon on the $i$-th reaction surface of standard deep crustal heating and extra neutrino crustal heating, respectively. $q_i$ is adopted from Table A.3 in Ref.~\cite{2008A&A...480..459H} where the initial compositions of the nuclear burning ashes are fixed to be $^{56}\rm Fe$. The total of neutrino heating is considered as ($0-1$) MeV per accreted nucleon for $1.4M_{\odot}$ neutron star, ($0-6$) MeV per accreted nucleon for $2M_{\odot}$ neutron star in present work.

%neutrino emission
We consider the bremsstrahlung of nucleon-nucleon and electron-ion, modified Urca, electron-positron pair, photo and plasmon processes for the slow cooling processes~\cite{2001PhR...354....1Y}. These neutrino emissivities are estimated approximately $10^{19-21}T_9^8~\rm erg~cm^{-3}~s^{-1}$, where $T_9$ is the local temperature in units of $10^9~\rm K$. For the fast cooling processes, the nucleon direct Urca process is considered. The corresponding neutrino emissivity is calculated approximately $10^{27}T_9^6~\rm erg~cm^{-3}~s^{-1}$~\cite{2001PhR...354....1Y}. As another fast cooling process, we consider the pion condensation process in the core, where the neutrino emissivity is given as around $10^{25}T_9^6~\rm{erg~cm^{-3}s^{-1}}$ according to Refs.~\cite{1993PThPS.112..221M,1994ApJ...431..309U}.

%For the low symmetry energy equation of state such as Togashi, whose direct Urca is forbidden for its low proton fraction~\cite{2019PTEP.2019k3E01D,2018IJMPE..2750067M}, we adopt a fast cooling due to pion condensation in the core~\cite{2021PhRvD.103f3009L}.

%gravitational energy
In the accretion layer, the mass fraction coordinate with changing mass $q=M_{r}/M(t)$ is adopted. It is appropriate to the computations of stellar structure with the changing stellar mass $M$~\cite{1981PThPS..70..115S}. As a result, the gravitational energy release $\varepsilon_{g}$ can be divided as~\cite{1984ApJ...278..813F}:
\begin{equation}
  \varepsilon_{g}^{(\rm nh)}=-{\rm exp} \left(-\frac{\phi}{c^2}\right)\left(T\frac{\partial s}{\partial t}\Bigg|_q+\mu_i\frac{\partial N_i}{\partial t}\Bigg|_q\right),
\label{eq:nh}
\end{equation}
\begin{equation}
  \varepsilon_{g}^{(\rm h)}={\rm exp} \left(-\frac{\phi}{c^2}\right)\frac{\dot{M}}{M}\left(T\frac{\partial s}{\partial {\rm ln}q}\Bigg|_t+\mu_i\frac{\partial N_i}{\partial {\rm ln}q}\Bigg|_t\right),
\label{eq:h}
\end{equation}
 where $\dot{M}_{-10}$ is mass-accretion rate in units of $10^{-10}~M_{\odot}~\rm yr^{-1}$, $s$ and $t$ are specific entropy and Schwarzschild time coordinate, respectively. $\mu_i$ and $N_i$ are chemical potential and number per unit mass of the $i-th$ elements, respectively. Eqs. (\ref{eq:nh}) and (\ref{eq:h}) are called nonhomologous and homologous terms, respectively. The latter indicates a homologous compression due to the accretion which causes the compressional heating. The nonhomologous term of the gravitational energy release will vanish in a long evolution time. To be consistent with the quiescent luminosity calculations as those of previous works which do not include the compressional heating, we turn off the compressional heating in our calculations.

 An outer most mesh-point is set at $q=1-9.9\times10^{-20}$, which is close enough to the photosphere, we observe the luminosity at the outmost mest-point as the closest thing to the total luminosity $L$. Then, the radiative zero boundary conditions can be written as follows:
 \begin{eqnarray}
P &=& \frac{GMM(t)\left(1-q\right)}{4\pi R^4}\left(1-\frac{2GM}{Rc^2}\right)^{-1/2}~, \label{eq:A1} \\
L &=& \frac{4\pi cGM}{\kappa}\frac{4aT^4}{3P}\frac{1+\frac{\partial \log\kappa}{\partial \log P}}{4-\frac{\partial \log\kappa}{\partial \log T}}\left(1-\frac{2GM}{Rc^2}\right)^{1/2}~. \label{eq:A2}
\end{eqnarray}

\begin{figure*}[htbp]
\begin{tabular}{cc}
\includegraphics[width=0.32\linewidth,angle=-90]{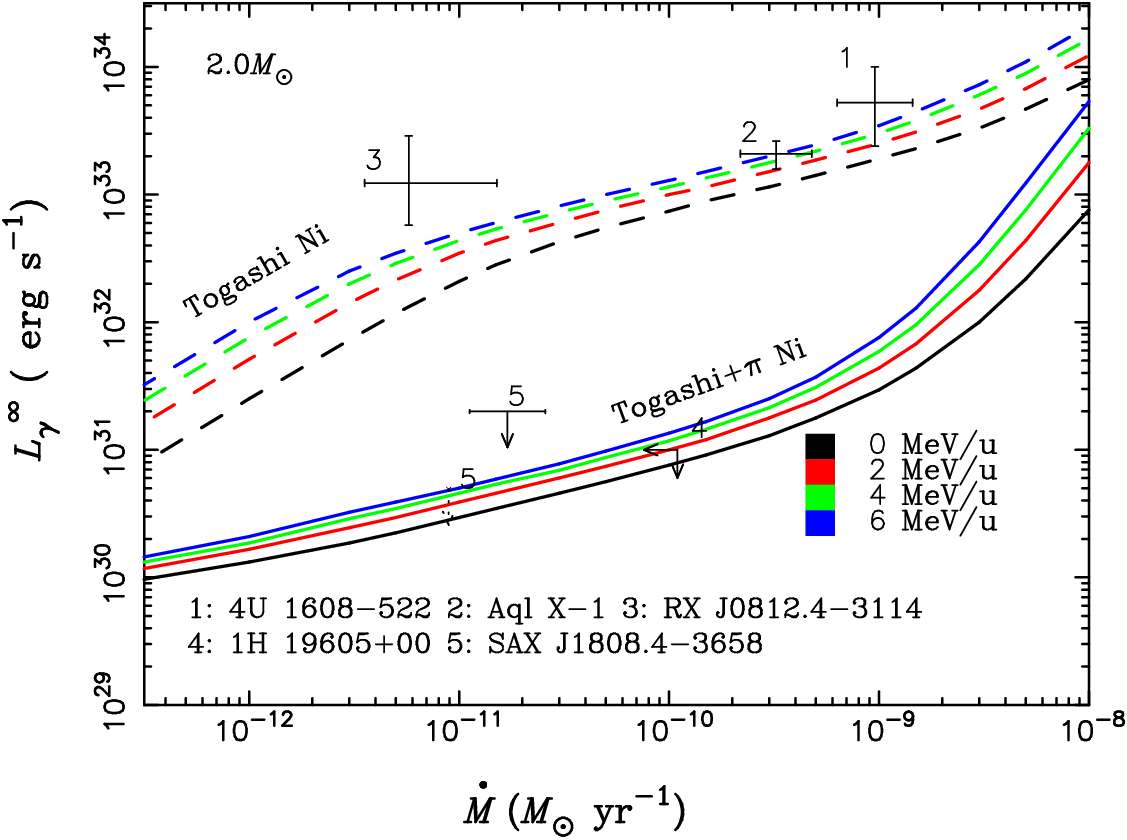}&
\includegraphics[width=0.32\linewidth,angle=-90]{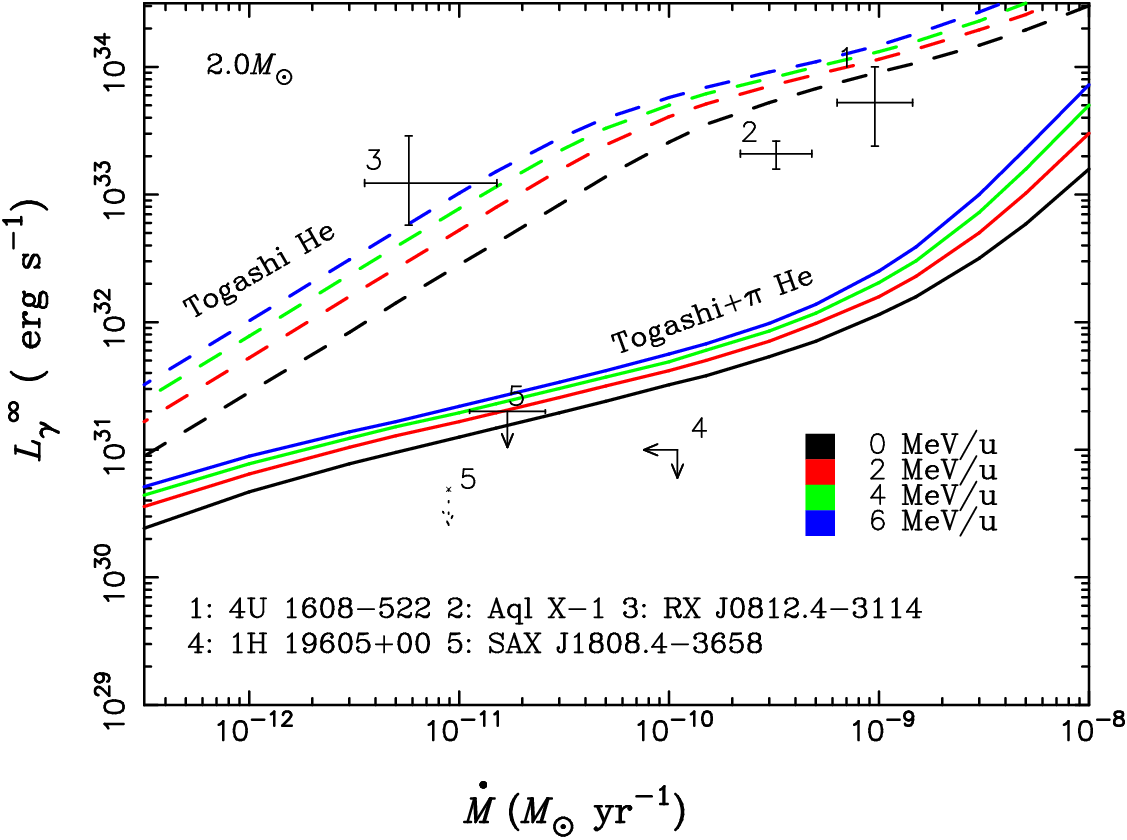}\\

\end{tabular}
    \caption{Quiescent luminosities of SXTs as functions of average accretion rates. The curves are compared with 5 representative observations which include three hottest sources: 1:4U 1608-522, 2:Aql X-1, 3:RX J0812.4-3114 and two coldest sources: 4:1H 19605+00, 5:SAX J180.4-3658 (Two error bars of source '5' represent the different spectral models, the solid error bar represents the highest estimate quiescent thermal luminosity while the dotted error bar represents the lowest estimate quiescent thermal luminosity by Heinke et al ~\cite{2009ApJ...691.1035H}), the five data are taken from Ref.~\cite{2019A&A...629A..88P}. Different colors represent the different values of neutrino heating. The dashed curves are produced with Togashi EOS, the solid curves are produced with Togashi+$\pi$ EOS.  The calculation models are for $2M_{\odot}$ neutron star. (a) With heavy elements surface as Ni (b) With light elements surface as pure helium
    %(b) The dashed curves are produced for a $1.4M_{\odot}$ neutron star while the dashed lines are produced for a $2M_{\odot}$ neutron star with the BSK24 EOS and pure helium surface.
    }
    \label{fig:lum}
\end{figure*}

We solve the above set of equations (\ref{eq:1})-(\ref{eq:6}) numerically with use of the Henyey-type numerical scheme of implicit method.

\section{Results}\label{sec:result}

In Table \ref{tab:nh1}, we show the redshifted quiescent luminosity of accreting neutron star with neutrino heating $q_{\nu}=0,2,4,6$ at different time average accretion rates for Togashi and BSK24 EOSs, respectively. The calculations are for $2M_{\odot}$ neutron star with the light elements surface as pure helium. We can see that the luminosities with neutrino heating $q_{\nu}=0,2,4,6$ increase obviously. Meanwhile, we show the redshifted quiescent luminosity of accreting neutron star with low neutrino heating $q_{\nu}=0,0.2,0.6,1.0$ at different time average accretion rates for $1.4M_{\odot}$ neutron star in Table \ref{tab:nh2}. It shows that luminosity is insensitive to the extra low neutrino heating, and the increment of the logarithm of redshifted quiescent luminosity with $q_{\nu}=1~\rm MeV/u$ is less than $0.2$ compared with the case $q_{\nu}=0$.

\begin{table*}[]
%\scriptsize
\centering
\caption{Logarithm of redshifted quiescent luminosity ($L_{\gamma}^{\infty}$ in units of $\rm erg ~s^{-1}$ ) of accreting neutron star with neutrino heating ($q_{\nu}$ in units of $\rm MeV/u$ ) at different time averaged accretion rates for Togashi and BSK EOSs, respectively. The calculations are for $2M_{\odot}$ neutron star with the light elements surface as pure helium.}
\label{tab:nh1}
\renewcommand\tabcolsep{6.5pt}
\begin{tabular}{ccccccccc}
\toprule
\multirow{2}{*}{$\dot{M}~(M_{\odot}~\rm yr^{-1})$}&\multicolumn{4}{c}{Togashi} &\multicolumn{4}{c}{BSK24}\\
\cmidrule(r){2-5}\cmidrule(r){6-9}
& $q_{\nu}=0$ & $2$ & $4$ & $6$
& $q_{\nu}=0$ & $2$ & $4$ & $6$\\
\midrule
$1.0\times10^{-8}$ & 34.48 & 34.62 &34.72 & 34.79 & 33.62&33.97 &34.24 &34.43\\
 $5.0\times10^{-9}$& 34.29& 34.41 &34.50 &34.57 &33.05 &33.35 &33.60 &33.82\\
$3.0\times10^{-9}$ & 34.17&34.29 &34.36 &34.43 &32.66 &32.92 &33.14 &33.33\\
$1.5\times10^{-9}$ & 34.13&34.15 &34.20 &34.26 &32.21 &32.43 &32.60 &32.75\\
$1.0\times10^{-9}$ & 33.96&34.06 &34.12 &34.17 &32.01 &32.20 &32.35 &32.47\\
$5.0\times10^{-10}$ &33.83 &33.94 &34.00 &34.04 &31.73 & 31.89&32.01 &32.10\\
$3.0\times10^{-10}$ &33.72 &33.85 &33.91 &33.96 &31.56 &31.72 & 33.82&31.90\\
$1.5\times10^{-10}$ &33.55 &33.71 &33.79 &33.84 &31.39 &31.53 &31.62 &31.68\\
$1.0\times10^{-10}$ & 33.41&33.61 &33.70 &33.76 &31.29 &31.43 &31.52 &31.58\\
$5.0\times10^{-11}$ & 33.14&33.39 &33.52 &33.60 &31.16 &31.28 &31.36 &31.42\\
$3.0\times10^{-11}$ & 32.92&33.19 &33.34 &33.44 &31.06 &31.19 &31.26 &31.31\\
$1.5\times10^{-11}$ & 32.62&32.90 &33.06 &33.18 &30.93 &31.05 &31.13 &31.18\\
$1.0\times10^{-11}$ & 32.45&32.72 &32.89 &33.01 &30.85 &30.98 &31.05 &31.11\\
$5.0\times10^{-12}$ & 32.15&32.42 &32.59 &32.71 & 30.71&30.85 &30.92 &30.98\\
$3.0\times10^{-12}$ &31.92 &32.20 &32.37 &32.49 &30.61 &30.75 &30.82 &30.88\\
$1.0\times10^{-12}$ & 31.45&31.72 &31.89 &32.01 &30.40 &30.54 &30.61 &30.67\\
$1.0\times10^{-13}$ &30.45 &30.72 &30.89 &31.01 &29.92 &30.07 &30.16 &30.22\\
\bottomrule
\end{tabular}
\end{table*}

\begin{figure*}[htbp]

\begin{tabular}{cc}
\includegraphics[width=0.32\linewidth,angle=-90]{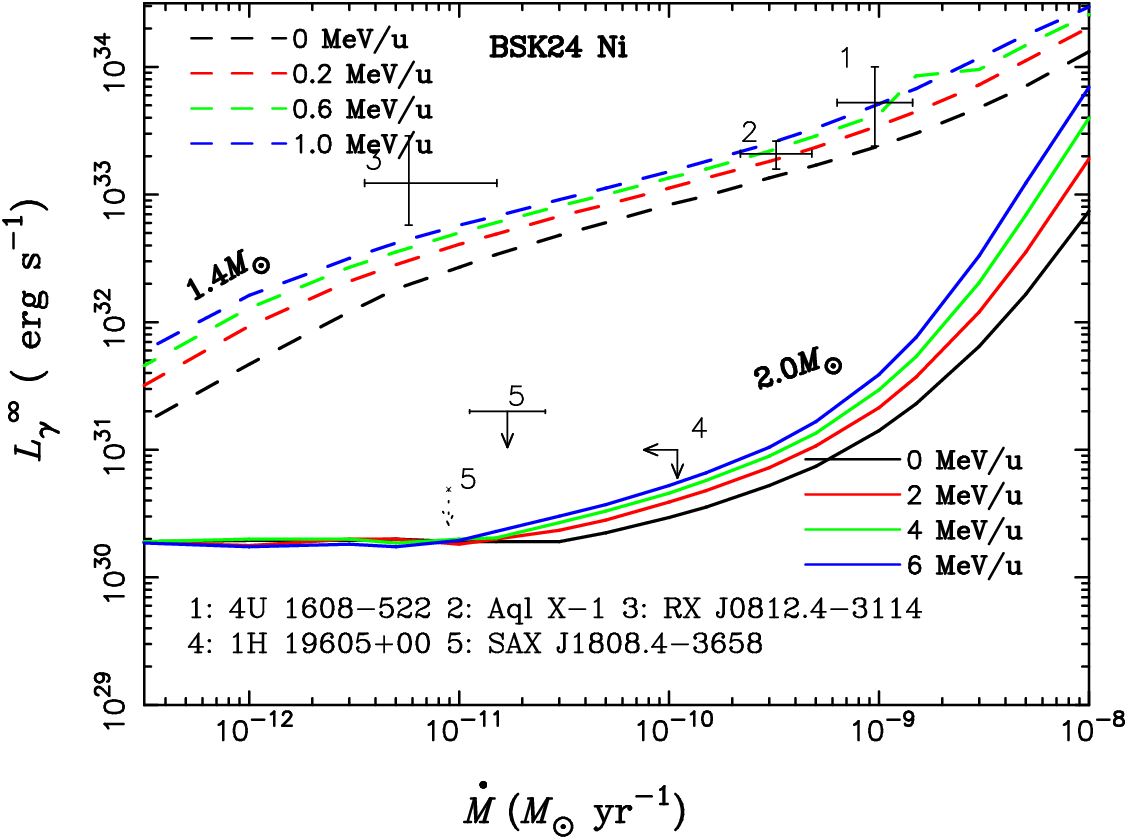}&
\includegraphics[width=0.32\linewidth,angle=-90]{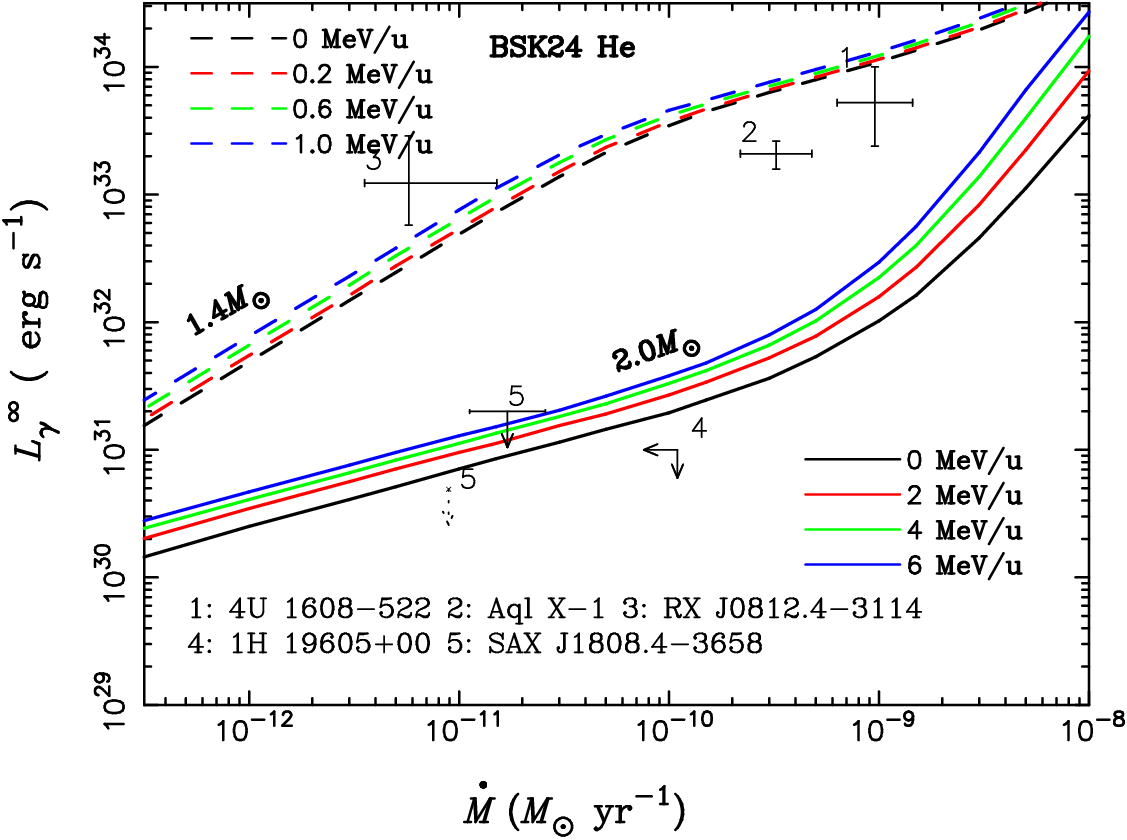}\\

\end{tabular}
   \caption{The same as Fig.~\ref{fig:lum} but with BSK24 EoS, the dashed lines are for $1.4M_{\odot}$ neutron stars and the solid lines are for $2.0M_{\odot}$ neutron stars. (a) With heavy elements surface as Ni. (b) With light elements surface as pure helium.}
   \label{fig:lumB}
\end{figure*}

\begin{table*}[]
%\scriptsize
\centering
\caption{The same as table \ref{tab:nh1} but for $1.4M_{\odot}$ neutron star and small neutrino heating.}
\label{tab:nh2}
\renewcommand\tabcolsep{6.5pt}
\begin{tabular}{ccccccccc}
\toprule
\multirow{2}{*}{$\dot{M}~(M_{\odot}~\rm yr^{-1})$}&\multicolumn{4}{c}{Togashi} &\multicolumn{4}{c}{BSK24}\\
\cmidrule(r){2-5}\cmidrule(r){6-9}
& $q_{\nu}=0$ & $0.2$ & $0.6$ & $1.0$
& $q_{\nu}=0$ & $0.2$ & $0.6$ & $1.0$\\
\midrule
$1.0\times10^{-8}$ & 34.66 & 34.68 &34.71 & 34.74 & 34.64&34.66 &34.70 &34.73\\
 $5.0\times10^{-9}$& 34.46& 34.47 &34.51 &34.53 &34.43 &34.45 &34.49 &34.52\\
$3.0\times10^{-9}$ & 34.32&34.34 &34.37 &34.40 &34.29 &34.31 &34.35 &33.38\\
$1.5\times10^{-9}$ & 34.03&34.14 &34.20 &34.30 &34.13 &34.15 &34.18 &34.21\\
$1.0\times10^{-9}$ & 34.08&34.10 &34.12 &34.15 &34.04 &34.06 &34.09 &34.12\\
$5.0\times10^{-10}$ &33.95 &33.96 &33.99 &34.01 &33.90 & 33.92&33.95 &33.98\\
$3.0\times10^{-10}$ &33.85 &33.87 &33.89 &33.92 &33.80 &33.82 & 33.85&33.88\\
$1.5\times10^{-10}$ &33.70 &33.72 &33.75 &33.78 &33.65 &33.67 &33.71 &33.74\\
$1.0\times10^{-10}$ & 33.59&33.61 &33.66 &33.69 &33.54 &33.57 &33.62 &33.66\\
$5.0\times10^{-11}$ & 33.37&33.40 &33.45 &33.50 &33.33 &33.37 &33.43 &33.47\\
$3.0\times10^{-11}$ & 33.17&33.21 &33.27 &33.32 &33.14 &33.18 &33.25 &33.31\\
$1.5\times10^{-11}$ & 32.88&32.92 &32.99 &33.04 &32.86 &32.90 &32.98 &33.05\\
$1.0\times10^{-11}$ & 32.71&32.75 &32.82 &32.87 &32.69 &32.73 &32.81 &32.88\\
$5.0\times10^{-12}$ & 32.41&32.45 &32.52 &32.58 & 32.39&32.44 &32.52 &32.58\\
$3.0\times10^{-12}$ &32.19 &32.23 &32.30 &32.36 &32.17 &32.21 &32.29 &32.36\\
$1.0\times10^{-12}$ & 31.71&31.75 &31.82 &31.88 &31.69 &31.74 &31.82 &31.89\\
$1.0\times10^{-13}$ &30.71 &30.75 &30.82 &30.88 &30.69 &30.74 &30.82 &30.89\\
\bottomrule
\end{tabular}
\end{table*}

In Fig.~\ref{fig:lum}, we examine the effects of neutrino heating on the quiescent luminosity of accreting neutron stars by comparing with the representative observations. For Togashi EOS, as the proton fraction is low that the direct Urca process is prohibited for any mass~\cite{2019PTEP.2019k3E01D}, it cools slowly even for a $2M_{\odot}$ neutron star (dashed lines in Fig.~\ref{fig:lum}). It is worth to note that the source RX J0812.4-3114 which has the low accretion rate $(5.82_{-2.29}^{+9.16}\times 10^{-12}~\rm{M_{\odot}~yr^{-1}})$ but high quiescent luminosity ($1.23_{-0.65}^{+1.66}\times 10^{33}~\rm{erg~s^{-1}}$) can be explained with the extra neutrino heating $q_{\nu}=4 ~\rm or~ 6$ MeV per accreted nucleon for a $2M_{\odot}$ neutron star with light elements surface. However, in the previous works, low mass neutron star with minimal cooling are proposed to explain the observation of RX J0812.4-3114~\cite{2019MNRAS.488.4427Z,2021PhRvD.103f3009L}.

On the other hand, some sources such as SAX J1808.4-3658 and 1H 1905+000 have low quiescent luminosity for high mass accretion rate, therefore, they are thought to be the strong evidence of enhanced neutrino cooling from neutron star. The pion condensation process is proposed for the low symmetry energy EoS~\cite{2021PhRvD.103f3009L,2021AD}. As a result, Togashi+$\pi$ EoS is also considered.
The quiescent luminosities for a $2M_{\odot}$ neutron star with Togashi+$\pi$ EoS are also shown in Fig.~\ref{fig:lum}.
The curves are indicated as the solid lines. We can see that the increment of luminosity due to neutrino heating is almost within the error bars of observation, a $2M_{\odot}$ neutron star with Ni surface and extra neutrino heating up to 6 MeV/u still can explain the coldest observations.

BSK24 EOS is also adopted for quiescent luminosity calculations in Fig.~\ref{fig:lumB} for a comparision. For $1.4M_{\odot}$ neutron star, as the direct Urca process is forbidden, it cools slowly. For the extra neutrino heating, the increment of luminosity is not obvious even with $q_{\nu}=1$ MeV per nucleon. It is similar for $1.4M_{\odot}$ neutron star with Togashi EOS in Table~\ref{tab:nh2}.  For a $2M_{\odot}$ neutron star, as the direct Urca process is switched on, it cools fast,
neutrino heating make a noticeable change in $L_{\gamma }^{\infty}-\langle\dot{M}\rangle$ cruves but it is still within the observation of luminosity error bars for $q_{\nu}=2-6$ MeV per accreted nucleon.

\begin{figure}[htbp]
    \centering
    \includegraphics[width=0.7\linewidth,angle=-90]{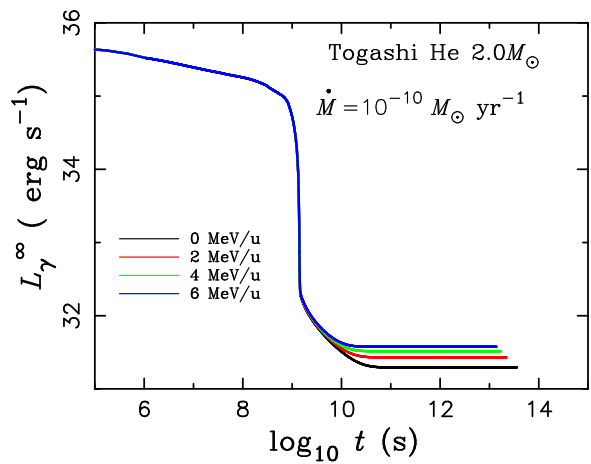}
    \caption{Time evolution of luminosities for a $2M_{\odot}$ neutron star with a light elements surface as pure helium until steady state. Different colors indicate different values of neutrino heating. The accretion rate is set at $\dot{M}=10^{-10}M_{\odot}~\rm yr^{-1}$.}
    \label{fig:LT}
\end{figure}

\begin{figure}[htbp]
    \centering
    \includegraphics[width=0.7\linewidth,angle=-90]{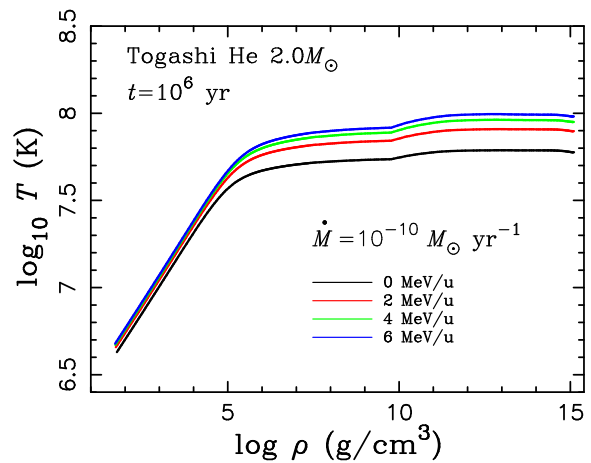}
    \caption{Temperature versus density at $t=10^6~\rm yr$ from the end of accretion, where the mass accretion rate is $\dot{M}=10^{-10}~\rm M_{\odot}~yr^{-1}$. Different colors represent different values of neutrino heating.}
    \label{fig:strc}
\end{figure}

To analyze the sensitivity of neutrino heating on the thermal evolution of accreting neutron stars, we display thermal evolution of a $2M_{\odot}$ neutron star with Togashi EOS and light elements surface as pure helium. As illustrated in Fig.~\ref{fig:LT}, the thermal evolution of the luminosity is begin sensitive to the neutrino heating at around $10^{10}~\rm s$. This is because the neutrino energy deposit the heat in the inner crust ($10^{12-13}~\rm g~cm^{-3}$). The luminosity in steady state increases as neutrino heating increases. For $t<10^{10}~\rm yr$, the luminosity is not affected by the neutrino heating. We illustrate the thermal structures of accreting neutron star with neutrino heating in Fig.~\ref{fig:strc}.
Notice that the behavior of the curves with different neutrino heating rate is similar with the standard deep crustal heating. That is, the temperature increases as neutrino heating increases, whose trend is more obvious in the high-density regions.

\section{Conclusion and outlook}\label{sec:conc}

In this paper, we have investigated the quiescent luminosity of transiently accreting neutron star with extra neutrino heating for the first time. As the strong gravity of neutron star will accelerate the in-falling particles to kinetic energies, the accreted matter (H or He) will undergo nuclear collisions with the nuclei on the surface of neutron star. For simplicity, we simulate the $\alpha-\rm Fe$ collision with incident beam energy 300-600 MeV/u by using the IBUU transport model, and get the charged pion production per collision event. The charged pion will decay and emit neutrinos. Assuming half of the neutrino carry their energy and move into the crust, we estimate the neutrino energy at different beam energies, and get $q_{\nu}=1.16 ~\rm MeV$ at the incident beam energy of 300 MeV per nucleon, $Q_{\nu}=6.55 ~\rm MeV$ at the beam energy of 600 MeV per nucleon. As the type of the target nuclei, electromagnetic acceleration as well as the parameters (e.g. impact parameter b, symmetry energy parameter x) in IBUU models would affect the pion production and the total neutrino energy. Here, we approximately assume that the neutrino heating due to the charged pion (both $\pi^{+}$ and $\pi^{-}$) decay deposit $0-1~\rm MeV$ per nucleon for $1.4M_{\odot}$ neutron star and $0-6~\rm MeV$ per nucleon for $2M_{\odot}$ neutron star in the inner crust.

We have found that the quiescent luminosity of $1.4M_{\odot}$ neutron star is insensitive to the small neutrino heating ($q_{\nu}=0-1~\rm MeV$ per nucleon). For $2M_{\odot}$ stars, the increment of luminosity is relatively obvious, but the coldest source can still be explained with Ni surface and $q_{\nu}=0-6~\rm MeV$ per nucleon. On the other hand, we have also found that the hot source such as RX J0812.4-3114 can be explained by a massive neutron star (e.g. $M\simeq2M_{\odot}$) with neutrino heating if the direct Urca process is prohibited. However, the enhanced cooling is still required to explain the low luminosities of coldest transiently accreting neutron stars. As the equation of state is still uncertain especially at high densities, one possibe mechanism to solve this apparent contradiction is that the codest observations can be fitted by a more massive neutron (e.g. $M>2M_{\odot}$) for which the direct Urca process turns on.
In a word, the quiescent luminosity of SXTs with extra neutrino heating is still consistent with the observations.

As a remark, the detailed calculation of neutrino energy deposition in the neutron star crust can be important for the shallow heating problem~\cite{2020PhRvC.102a5804C}. If the neutrino energy absorbed in the outer layers, it is possibly as the shallow heating source, and the different values of shallow heating can be caused by the different mass and surface composition of neutron star which leads to the different kinetic energy of in-falling particle and the target nuclei of collision. As we choose two representative neutron star masses as $M=1.4, 2.0M_{\odot}$ in this work, it is also worthwhile to investigate the neutrino energy deposition in the crusts of neutron stars with more different masses. Besides, we adopt Ref.~\cite{2008A&A...480..459H} as the traditional deep crustal heating input, an improved deep crustal heating model by considering the diffusion of neutrons in the inner crust is proposed~\cite{2020MNRAS.495L..32C,2021PhRvD.103L1301G,2021MNRAS.507.3860S}. It's also interesting to investigate the quiescent luminosity of SXTs with this modern deep crustal heating. We will leave these issues in near future. Moreover, as the experimental measurements of pion production cross sections in $p -\rm Fe$, $\alpha -\rm Fe$ collision are still incomplete, the comparison between observation and theory are useful for ongoing experiment program to measure $p -{\rm Fe}$, $\alpha -\rm Fe$ collisions. This paper provide a basis for these forthcoming studies.

\begin{acknowledgments}
We would like to thank the anonymous referee for detailed and constructive comments that helped to improve the paper.
This work has been supported financially by the National Natural Science Foundation of China under grant No. 11803026 and Xinjiang Natural Science Foundation under grant No. 2020D01C063. ZGD was supported by the National Key Research and Development Program of China (grant No. 2017YFA0402600), the National SKA Program of China (grant No. 2020SKA0120300), and the National Natural Science Foundation of China (grant No. 11833003). A.D. is partially supported by RIKEN iTHEMS Program.
\end{acknowledgments}

%\appendix

%\section{Appendixes}

%\nocite{*}

%\bibliography{ref}{}% Produces the bibliography via BibTeX.
%\bibliographystyle{apsrev4-2}

\end{document}

%% file: Name_Journal.tex
%===== Name of Journals =================================
%%% ApJ, ApJS, ApJL
%\newcommand{\apj}{Astrophys. J. }
\newcommand{\apjs}{Astrophys. J. Suppl. }
\newcommand{\apjl}{Astrophys. J. Lett. }
%%% PASJ
\newcommand{\pasj}{Publ. Astron. Soc. Japan. }
%%% PASA
\newcommand{\pasa}{Publ. Astron. Soc. Australia. }
%%% Physics Report
\newcommand{\physrep}{Phys. Rep. }
%PTP, PTEP
\newcommand{\ptp}{Prog. Theor. Phys. }
\newcommand{\ptps}{Prog. Theor. Phys. Suppl. }
\newcommand{\ptep}{Prog. Theor. Exp. Phys. }
%%% AIP Confference Proceedings
\newcommand{\AIP}{AIP Conf. Proc. }
%%% A & A
\newcommand{\aap}{Astron. Astrophys. }
%%% Space Sci. Rev.
\newcommand{\ssr}{Space Sci. Rev. }
%%% Nature
%\newcommand{\nat}{Nature }
%%% Science
\newcommand{\sci}{Science }
%%% Phys. Rev. C, D
%\newcommand{\prc}{Phys. Rev. C}
%\newcommand{\prd}{Phys. Rev. D}
%%%Phys. Rev. Lett
%\newcommand{\prl}{Phys. Rev. Lett. }
%%% New Astron. Rev.
\newcommand{\nar}{New Astron. Rev. }

\newcommand{\araa}{Ann. Rev. Astron. Astrophy. }
\newcommand{\mnras}{Mon. Not. Roy. Astron. Soc. }

%%%Nucl. Phys. A
\newcommand{\nphysa}{Nucl. Phys. A}

%JCAP
\newcommand{\jcap}{JCAP}
%Memorie della Società Astronomica Italiana

\newcommand{\memsai}{Memorie della Soc. Astron. Ital.}